\documentclass[preprint,showpacs,showkeys,aps,prd,amsfonts,eqsecnum,nofootinbib,superscriptaddress]
{revtex4} 
\usepackage{graphicx,epsfig}
\begin{document}
%
\preprint{CUMQ/HEP 132, WM-04-122}
%
%
\title{Neutrino Masses in Effective Rank-5 Subgroups of $E_6$ I: Non-Supersymmetric Case}
\author{Mariana Frank}
\email{mfrank@vax2.concordia.ca}
\affiliation{Department of Physics, Concordia University, 7141 Sherbrooke St.
West, Montreal, Quebec, CANADA H4B 1R6}
\author{Marc Sher}
\email{sher@physics.wm.edu}
\affiliation{Particle Theory Group, Department of Physics,
College of William and Mary, Williamsburg, VA 23187-8795, USA}
\author{Ismail Turan}
\email{ituran@physics.concordia.ca}
\affiliation{Department of Physics, Concordia University, 7141 Sherbrooke St.
West, Montreal, Quebec, CANADA H4B 1R6}
\date{\today}

\begin{abstract}
The neutral fermion sectors of $E_6$-inspired low energy models, in 
particular the Alternative Left-Right and Inert models,
are considered in detail within the non-supersymmetric scenario. We 
show that in their simplest form, these models always predict, for each generation, the lightest neutrino to be an 
$SU(2)_L$ singlet, as well as two extra neutrinos with masses of the 
order of the up-quark mass.  In order to recover Standard Model 
phenomenology, additional assumptions in the form of  discrete 
symmetries and/or  new interactions are needed. These are
classified as the Discrete Symmetry (DS),  Higher Dimensional 
Operators (HDO), and Additional Neutral Fermion (ANF) methods. 
The DS method can solve the problem, 
but requires additional Higgs doublets that do not get vacuum 
expectation values.  The HDO method 
predicts no sterile neutrino, and that the active neutrinos mix 
with a heavy isodoublet neutrino, thus   
slightly suppressing the couplings of active neutrinos, 
with interesting phenomenological implications.  
The ANF method also predicts this suppression, 
and also naturally includes one or more ``sterile" 
neutrinos. This scenario allows the existence of sterile neutrino(s) 
in either a   
${\bf 3}+{\bf 1}$ or ${\bf 2}+ {\bf 2}$ structure at low energies, which are favored by the LSND
result.
\end{abstract}
\vskip -2.5cm
\maketitle
\section{Introduction}\label{sec:intro}

The discovery of 
solar \cite{solar} and atmospheric \cite{atm} neutrino oscillations has provided the first 
confirmed scenario of physics  beyond the Standard Model. The 
combined results from solar, atmospheric and long baseline neutrino 
experiments are well described by oscillations of three active 
neutrinos $\nu_e,~\nu_{\mu}$ and $~\nu_{\tau}$, with mass squared 
splittings estimated to be $5.4 \times 10^{-5}< \Delta m_{sol}^2< 9.5 
\times 10^{-5}$ eV$^2$  and $1.2 \times 10^{-3}< \Delta m_{atm}^2 < 
4.8 \times 10^{-3}$ eV$^2$ \cite{splitting}. However,  the Los Alamos Liquid 
Scintillation Detector (LSND) requires $10 > \Delta m^2 > 0.2 $ eV$^2$ \cite{LSND}, 
a serious disagreement with the other results. The MiniBooNE 
experiment at Fermilab \cite{miniboone} is in the process of checking the 
validity of the LSND experiment.  Taking at face value the LSND 
results, a minimum of four neutrinos seems to be required to explain 
all available neutrino data.  LEP-SLC measurements of the $Z$ decay 
width restrict the number of active neutrinos to three; thus one or 
more of the neutrinos must be ``sterile" \cite{sterile}. 
Such scenarios have been 
studied extensively \cite{Sorel:2003hf, Mohapatra:2004uy, Babu:2004mj, Godfrey:2004gv, Krolikowski:2004ru, McDonald:2004pa, 
Barger:2003xm, Stephenson:2003ta, Paes:2002ah, Donini:2001qv}. Mixing of sterile and active neutrinos affects 
directly the present neutrino experiments and limits have been set on 
such mixings.  A valid question remains: how natural is it, in a 
beyond the Standard Model scenario, to obtain physically 
acceptable mixings between sterile and active neutrinos, while 
maintaining the constraints from weak scale phenomenology.

Several 
extensions of the Standard Model predict the existence of exotic 
fermions. Of these, superstring theories represent the most promising scenario 
for a unified theory of all fundamental interactions. One set of superstring theories are anomaly-free ten dimensional 
theories based on $E_8 \times E_8$ heterotic strings coupled to N=1 
gravity \cite{superstrings}, with matter belonging to the {\bf 27} representation of $E_6$. Previous interest in the $E_6$ GUTs dates as far as 1970's \cite{E6}
when it was noted that $E_6$ was the next anomaly-free choice group 
after $SO(10)$, and that each generation of fermions can be placed in 
the {\bf 27}-plet representation. 

The $E_6$ spectrum contains several 
neutral exotic fermions, some which could be interpreted as sterile 
neutrinos. The precise details of mass generation and mixing with the 
active neutrinos would depend on which subgroup of $E_6$ is 
considered.  There are many phenomenologically acceptable low energy 
models which arise from $E_6$. In this work we concentrate on rank-5 
subgroups, which can always break to $SU(3)_C \times SU(2)_L \times 
U(1)_Y \times U(1)_{\eta}$ \cite{HR,Langacker}.

We analyze neutrino masses and mixings, 
as well as active-sterile neutrino assignments and mixing in group 
decompositions of $E_6$ under the maximal subgroup $SU(3)_C 
\times SU(3)_L \times SU(3)_H$ to the Standard Model. These intermediate subgroups can include extra 
$SU(2)$ groups, which give rise to the usual Left-Right symmetric model (LR) \cite{LR}, the 
Alternative Left-Right symmetric model (ALR) \cite {ALR} and the Inert model \cite{HR,inert}. Though there are small 
differences among these groups with regards to neutrino masses and 
mixing, we shall be able to present a study applicable to all. We 
keep this discussion valid for the non-supersymmetric case, leaving 
the details for the supersymmetric scenario to another work \cite{susy}. 

Our 
paper is organized as follows. We discuss these models in Section 2. 
In sections 3 and 4 we analyze neutrino masses and mixings in the 
Alternative Left-Right and Inert models, respectively. Both of these 
models suffer from predicting too large  a Dirac mass for the active 
neutrinos. We suggests mechanisms to rectify this problem in Section 
5. We discuss the implications of our results and conclude in Section 6.

\section{The Models}\label{sec:model}

The fundamental representation of $E_6$, the $\bf{27}$-plet, branches into 
\begin{eqnarray}
\bf{27}&=&(3^c,3,1)+(\bar{3}^c,1,\bar{3})+(1^c,\bar{3},3)\nonumber\\
&=&\;\;\;\;\;\;q\;\;\;\;\;+\;\;\;\;\; \bar{q}\;\;\;\;\;\;+\;\;\;\;\;l
\label{333}
\end{eqnarray}
under the maximal subgroup, $SU(3)_C\otimes SU(3)_L\otimes
SU(3)_H$.   The particle content of the $\bf{27}$-plet for one family
under this decomposition can be written as
\begin{eqnarray}
q=\left(\begin{array}{c}
u \\
d \\
h
\end{array}
\right)_L,\;\;
\bar{q}=\left(u^c\;d^c\;h^c\right)_L,\;\;
l=\left(\begin{array}{ccc}
E^c & N & \nu \\
N^c & E   & e \\
e^c & \nu^c & S^c
\end{array}\right)_L.
\end{eqnarray}
Here we have used the notation that $SU(3)_L(SU(3)_H)$ operates vertically
(horizontally) and the minus signs in front of the fields are
suppressed.\footnote{We write fields as left-chiral Dirac spinors and
throughout the rest of the paper we use $f_L^c$ for a fermion field $f$ as
a shorthand notation for $(f^c)_L$, as we know that the chiral projection
and conjugation do not commute.  Thus, $f_L^c\equiv (f_R)^c=C \gamma^0
f_R^*$ where $C=\left( \begin{array}{cc}-\epsilon&0\\0 & \epsilon\\
\end{array}\right)$.  Here we adopt the chiral representation and
$\epsilon\equiv i\sigma_2$.}  There are three ways to embed an $SU(2)_H$ 
into the $SU(3)_H$, just as $I$-spin, $U$-spin and $V$-spin can be 
embedded in the $SU(3)$ flavor group.
The best-known breaking is when the first and the second columns form a
$SU(2)_H$ doublet; this corresponds to the LR symmetric model ($H\!=\!R$).  An 
alternative version is when the
first and the third columns form an $SU(2)_H$ doublet; this corresponds to 
the ALR symmetric model ($H\!=\!R^{\prime}$).  Finally,  the second and the third 
columns can form an $SU(2)_H$ doublet; this corresponds to the Inert
model ($H\!=\!I$).  In LR, $\left(u^c\; d^c\right)_L(\left(e^c\;\nu^c\right)_L),
\left(\begin{array}{cc} E^c & N \\ N^c & E \end{array}\right)_L$, and
$h_L^c$ (and the third column of $l$) become $SU(2)_R$ doublets,
a bi-doublet, and singlets, respectively.  For the ALR case, $\left(h^c\;
u^c\right)_L(\left(e^c\;S^c\right)_L), \left(\begin{array}{cc} E^c & \nu \\
N^c & e \end{array}\right)_L$, and $d_L^c$ (and the particles in the second
column of $l$) are the corresponding ones under $SU(2)_{R^{\prime}}$. 
Finally in the Inert model, $\left(h^c\; d^c\right)_L(\left(\nu^c\;S^c\right)_L),
\left(\begin{array}{cc} N & \nu \\ E & e \end{array}\right)_L$, and
$u_L^c$ (and the particles in the first column of $l$) are the corresponding
multiplets under $SU(2)_I$. 

To determine the $U(1)$ quantum numbers, we need to look at the 
electromagnetic charge operator.  If we consider the case where only $SU(3)_L$ is broken
down to $SU(2)_L\otimes U(1)_{Y_L}$, the electromagnetic charge
$Q_{em}=I_{3L}+Y/2$ for all $\bar{q}$ becomes zero.  Therefore, $SU(3)_H\to
SU(2)_H\otimes U(1)_{Y_H}$ is needed such that $SU(2)_H$ and/or
$U(1)_{Y_H}$ can contribute to $Q_{em}$.   When
both $SU(2)_H$ and $U(1)_{Y_H}$ contribute to $Q_{em}$, we end up with the
LR \footnote{This is the rank-6 version of the familiar
LR symmetric model.} and ALR symmetric models.  The
``Inert'' model, is obtained when the $SU(2)_H$ does not contribute to
$Q_{em}$.  We will use the notation $H=R,R^{\prime},I;
Y_H=Y_{R,R^{\prime},I}$ for the LR, ALR and Inert groups, respectively. 
The gauge groups are at this level $SU(3)_C\otimes SU(2)_L\otimes
SU(2)_R\otimes U(1)_L\otimes U(1)_R,\,SU(3)_C\otimes SU(2)_L\otimes
SU(2)_{R^{\prime}}\otimes U(1)_L\otimes U(1)_{R^{\prime}}$, and
$SU(3)_C\otimes SU(2)_L\otimes SU(2)_I\otimes U(1)_Y\otimes U(1)^{\prime}$
for LR, ALR and Inert cases, respectively \cite{HR, Harada:2003sb}.  It is further possible to break
them into some effective rank-5 forms by reducing $U(1)_L\otimes
U(1)_{R(R^{\prime})}\to U(1)_{V=L+R(R^{\prime})}$ for the LR (ALR) case and
$SU(2)_I\otimes U(1)^{\prime}\to SU(2)_I$ for the Inert case.
 The quantum numbers of the particles in ALR and
Inert models are given in Table \ref{tablepartic}.  
\begin{table}[htb]
	\caption{The quantum numbers of fermions in $\bf{27}$ of $E_6$ at $SU(3)_C\otimes SU(2)_L\otimes SU(2)_{R^{\prime}}\otimes U(1)_{V=Y_L+Y_{R^{\prime}}}$ and $SU(3)_C\otimes SU(2)_L\otimes SU(2)_I\otimes U(1)_Y$ levels.} \label{tablepartic}     
\begin{center}
    \begin{tabular}{ccccccc}
    \hline\hline
\vspace{0.1cm}
$\;\;\;\;\;$ state $\;\;\;\;\;$ & $\;\;\;\;\;I_{3L}\;\;\;\;\;$ & $\;\;\;\;\;I_{3R^{\prime}}\;\;\;\;\;$ & $\;\;\;\;\;I_{3I}\;\;\;\;\;$ & 
$\;\;\;\;\;V/2\;\;\;\;\;$ & 
$\;\;\;\;\;Y/2\;\;\;\;\;$ & $\;\;\;\;\;Q_{em}\;\;\;\;\;$\\ 
\hline 
    $u_L$      &  1/2    &  0     &  0     &  1/6   & 1/6   &  2/3     \\
    $u_L^c$    &  0      &  -1/2   &  0     &  -1/6  & -2/3  &  -2/3     \\
    $d_L$      &  -1/2   &  0     &  0     &  1/6   &  1/6  &  -1/3     \\
    $d_L^c$    &  0      &  0     &  -1/2   &  1/3   &  1/3  &  1/3    \\
    $h_L$      &  0      &  0     &  0     &  -1/3  &  -1/3 &  -1/3     \\
    $h_L^c$    &  0      &  1/2  &  1/2  &  -1/6   &  1/3  &  1/3     \\
    $e_L$      &  -1/2   &  -1/2  &  -1/2  &  0     &  -1/2 &  -1     \\
    $e_L^c$    &  0      &  1/2   &  0     &  1/2   &  1    &  1     \\
    $E_L$      &  -1/2   &  0     &  1/2   &  -1/2  &  -1/2 &  -1     \\
    $E_L^c$    &  1/2    &  1/2   &  0     &  0     &  1/2  &  1     \\
    $\nu_L$    &  1/2    &  -1/2  &  -1/2  &  0     &  -1/2 &  0     \\
    $\nu_L^c$  &  0      &  0     &  1/2   &  0     &  0    &  0     \\
    $N_L$      &  1/2    &  0     &  1/2   &  -1/2  &  -1/2 &  0     \\
    $N_L^c$    &  -1/2   &  1/2   &  0     &  0     &  1/2  &  0     \\
    $S_L^c$    &  0      &  -1/2  &  -1/2  &  1/2   &  0    &  0     \\
\hline \hline
        \end{tabular}
        \end{center}
\vskip -0.5cm
\end{table}

The Higgs sector of the model is sometimes found by assuming, in the spirit of 
 SUSY models, that the allowed representations also come from a ${\bf 27}$-plet. 
However, since we are not considering SUSY models, we do not assume that all of the states in the ${\bf 27}$-plet 
are present (so colored scalars will not be introduced, for example).
For the ALR model, we can have  $H_S$, singlet under both $SU(2)$ groups, 
$H_1$ doublet under $SU(2)_{R^{\prime}}$ and singlet under $SU(2)_L$, 
$H_2$ doublet under $SU(2)_L$ and singlet under $SU(2)_{R^{\prime}}$, and a
bi-doublet $H_3$.  The neutral components of $H_S, H_1, H_2$, and $H_3$ are scalars with the same quantum 
numbers as $\nu_L^c, S_L^c, 
N_L$, and $(N_L^c,\nu_L)$ and they are from
$(\bf{16},\bf{1}), (\bf{1},\bf{1}), (\bf{10},\bf{\overline{5}})$, and
$((\bf{10},\bf{\overline{5}}),(\bf{16},\bf{\overline{5}}))$ representations under
$(SU(10),SU(5))$, respectively.  In the case of the Inert model, however, 
the representations are
slightly different \cite{inert}.  There is no singlet scalar field $(H_S)$ under
$SU(2)_I$ but an additional neutral $SU(2)_I$ doublet $H_D$ is
needed.  This doublet corresponds to the components $\nu_L^c$ and $S_L^c$
of the fermion doublet. We parametrize these Higgs doublets vev's as 
\begin{eqnarray}
\langle H_1 \rangle = \left(0\;N_1\right),\;\;\langle H_2 \rangle =  \left(\begin{array}{c}
v_1\\
0 \\
\end{array}
\right),\;\; \langle H_3 \rangle =\left(\begin{array}{cc}
0&v_2\\
v_3&0 \\
\end{array}
\right),\;\;\langle H_S\rangle=N_2,
\label{higgsALR}
\end{eqnarray}
in the ALR model and  
\begin{eqnarray}
\langle H_D \rangle = \left(N_2\;N_1\right),\;\;\langle H_2 \rangle =  \left(\begin{array}{c}
0\\
v_3 \\
\end{array}
\right),\;\; \langle H_3 \rangle =\left(\begin{array}{cc}
v_1&v_2\\
0&0 \\
\end{array}
\right),
\end{eqnarray}
in the Inert model.  The quantum numbers and vev's of the color-singlet,
neutral Higgs fields in $\bf{27}$ of $E_6$ are given in Table
\ref{tablehiggs}.  
\begin{table}[htb] 
\caption{The quantum
numbers of fermions in $\bf{27}$ of $E_6$ at $SU(3)_C\otimes SU(2)_L\otimes
SU(2)_{R^{\prime}}\otimes U(1)_{V=Y_L+Y_{R^{\prime}}}$ and $SU(3)_C\otimes
SU(2)_L\otimes SU(2)_I\otimes U(1)_Y$ levels.} \label{tablehiggs}
\begin{center} 
\begin{tabular}{cccccc}
    \hline\hline
\vspace{0.1cm}
 $\;\;\;\;\;$vev$\;\;\;\;\;$ & $\;\;\;\;\;I_{3L}\;\;\;\;\;$ & $\;\;\;\;\;I_{3R^{\prime}}\;\;\;\;\;$ & 
$\;\;\;\;\;I_{3I}\;\;\;\;\;$ & $\;\;\;\;\;V/2\;\;\;\;\;$ & $\;\;\;\;\;Y/2\;\;\;\;\;$ \\ 
\hline 
   $v_1$    &  1/2    &  0     &  1/2   &  -1/2 &  -1/2   \\
    $v_2$    &  1/2    &  -1/2  &  -1/2  &  0    &   -1/2  \\
    $v_3$    &  -1/2   &  1/2   &  0     &  0    &   1/2    \\
    $N_1$    &  0      &  -1/2  &  -1/2  &  1/2  &   0     \\
    $N_2$    &  0      &  0     &  1/2   &  0    &   0     \\
\hline \hline
        \end{tabular}
        \end{center}
\end{table}
We assume that the $SU(2)_L$ doublets acquire vev's $v_i\sim 10^2$
GeV, the symmetry breaking scale of the electroweak gauge group, while the
$SU(2)_L$ Higgs singlets get vev's $N_i$ much larger than the
scale of the electroweak symmetry breaking (that is, $N_i\gg v_i$).  This hierarchy is
needed from the fact that no experimental signal 
for the exotic quarks and leptons has been observed.  The mass terms for
the fermions can be obtained from the dimension-4 Yukawa interactions of
the form ${\cal L}_Y=\lambda \psi({\bf{27}})\psi({\bf{27}})H(\bf{27})$.  Here
$\psi(\bf{27})$ is the $\bf{27}$-plet of $E_6$ involving leptons and
quarks, and $H(\bf{27})$ is the one involving Higgs scalars.
  The coefficient $\lambda$ represents the corresponding generation
dependent Yukawa coupling, where generation indices are suppressed. 
The explicit mass terms in the above Lagrangian ${\cal L}_Y$ can be written
using the fact that each term has total hypercharge $Y$ zero and is
invariant under the gauge group of the model under consideration (that is,
terms invariant under the $SU(3)_C\otimes SU(2)_L\otimes
SU(2)_{R^{\prime}}\otimes U(1)_V$ gauge group for the ALR model and under
the $SU(3)_C\otimes SU(2)_L\otimes SU(2)_I\otimes U(1)_Y$ gauge group for
the Inert model).  Therefore, all the allowed Yukawa terms can be written
with the use of the Tables \ref{tablepartic} and \ref{tablehiggs}.  We 
consider the neutral sector of the $\bf{27}$-plet of $E_6$ in
the rest of the paper for the ALR and Inert models. Similar results can be obtained for LR models.  

\section{Neutrinos in the ALR Symmetric Model}\label{sec:ALR}
We now look at the allowed Yukawa couplings in the ALR model.  For
convenience, we use the following notation: 
\begin{eqnarray}
Q&=&\left(\begin{array}{c}u\\d\end{array}\right)_L(3,2,1,1/6)\,,\;X^c=\left(h^c\,\,u^c\right)_L(\overline{3},1,2,-1/6)\,,\;L^{\prime}=\left(\begin{array}{c}N\\E\end{array}\right)_L(1,2,1,0)\,,\nonumber\\
F&=&\left(\begin{array}{cc} E^c & \nu \\ N^c & e
\end{array}\right)_L(1,2,2,0)\,,\;L^c=\left(e^c\,\,S^c\right)_L(1,1,2,1/2)\,.
\end{eqnarray}
Then, all possible Yukawa terms which are $SU(3)_C\otimes SU(2)_L\otimes
SU(2)_{R^{\prime}}\otimes U(1)$ invariant can be written using of the
Higgs fields in Eq.~(\ref{higgsALR}).  The Yukawa Lagrangian is 
\begin{eqnarray} 
{\cal L_Y}&=& -\lambda_1\left[L^cFH_2+L^cH_3L^{\prime}+H_1FL^{\prime}
\right]+\frac{\lambda_2}{2}\left[FFH_S+FH_3\nu_L^c\right]+\lambda_3QH_3X^c\nonumber\\
&&+\lambda_4 d_L^c Q H_2+\lambda_5 h_L X^c H_1+\lambda_6 h_Ld^c_LH_S\,,
\label{YukwALR}
\end{eqnarray}
where we suppress all generation indices and use a shorthand notation 
for each term.  So, for example, $L^cFH_2$ should be read as
$(L^c)^T\,\epsilon\,F\,\epsilon\,H_2$ with $\epsilon=i \sigma_2$. 
 The part of the Lagrangian relevant to our
discussion here is (when the Higgs fields get vev's) 
\begin{eqnarray} 
{\cal L_Y^{\rm 0}}&=&\lambda_1\left[v_1(e_Le_L^c-N_L^c
S_L^c)-v_2 e_L^c E_L-v_3 N_L S_L^c+N_1(E_L E_L^c-N_L
N_L^c)\right]\nonumber\\ &&+\lambda_2\left[v_2 \nu_L^c N_L^c + v_3 \nu_L
\nu_L^c+N_2(- e_L E_L^c+\nu_L N_L^c)\right]+\lambda_3 v_3 u_Lu^c_L\,,
\label{ALRyukawa} 
\end{eqnarray}
where we have suppressed family indices and include charged lepton terms  and part of
the $\lambda_3$ term for later convenience.\footnote{Since this paper is concentrating 
on neutrinos, we will not discuss mixing between light and heavy fields in the charged 
lepton or quark sectors.  Such mixing can have a wide 
range of interesting phenomenological effects, see Ref. \cite{Frampton:1999xi} 
for a detailed discussion and a 
list of references.} Here 
it should be understood
that the $e_Le_L^c$ term, for example, stands for $(e^c)_L^T\,C\,e_L\equiv
\overline{e_R}e_L$.

From the above Yukawa interactions, the Majorana mass matrix for the neutral fields in the
 $\left(\nu_L,N_L,N_L^c,\nu_L^c,S_L^c\right)$ basis becomes (for one generation)
\begin{eqnarray}
{\cal M}_{\rm neutral}= \left(\begin{array}{ccccc}
\vspace{0.18cm}
       0      &      0         & \lambda_2 N_2  & \lambda_2 v_3  &     0          \\
\vspace{0.18cm}
       0      &      0         & -\lambda_1 N_1  & 0              & -\lambda_1 v_3  \\
\vspace{0.18cm}
\lambda_2 N_2 &  -\lambda_1 N_1 &       0        & \lambda_2 v_2  & -\lambda_1 v_1  \\
\vspace{0.18cm}
\lambda_2 v_3 &      0         & \lambda_2 v_2  &       0        &     0          \\ 
\vspace{0.18cm}
      0      &  -\lambda_1 v_3 & -\lambda_1 v_1  &       0        &     0          \\ 
\end{array}
\right)\,.\label{MatrixALR}
\end{eqnarray}
Further we define $\lambda_1 v_1\equiv m_{ee^c}, \lambda_1 N_1\equiv
m_{EE^c}$, and $\lambda_2 v_3\equiv m_{\nu\nu^c}$ since it is clear from
 Eq.~(\ref{ALRyukawa}) that $m_{ee^c}, m_{EE^c}$, and $m_{\nu\nu^c}$ are
the Dirac mass terms for the electron $e_L$, the exotic charged lepton
$E_L$, and the ordinary (active) neutrino $\nu_L$.  Note that the SM
(active) neutrino gets Dirac mass from the same source as the up quark.  
Thus, at the first stage, there appears to be a large Dirac mass problem for the
neutrinos unless there is an (unnatural) hierarchy 
$\lambda_2\ll \lambda_3$.  Unlike the ``conventional" see-saw model, we do 
not have a large Majorana mass term for the right-handed neutrino, so 
other techniques must be used to deal with this large mass.  
This problem is also severe in both
the Inert and the ordinary LR symmetric models where the active neutrinos
and up quark (the electron for LR case) get their Dirac masses from the same source. 
We will discuss the Inert model case in the next section.  For the
ordinary LR symmetric model, see \cite{Mohapatra:1987nx, Nandi:1985uh} for further details.

The secular equation for the eigenvalues cannot be solved exactly, and so 
we expand in powers of $v_i/N_i$.   In this approximation (neglecting $O(v_i/N_i)$ terms), there are two roots of
the secular equation which correspond to states with mass eigenvalue $\pm
m_{\nu\nu^c}$. The other three mass eigenvalues can also be determined, again under the 
assumption that $\lambda_2v_2\sim m_{\nu\nu^c}\sim m_{ee^c} \ll 
\lambda_2N_2\sim m_{EE^c}$
\begin{eqnarray} 
R_1&\simeq&-\frac{2
m_{\nu\nu^c}\left(m_{ee^c}m_{EE^c}+\lambda_2^2 v_2
N_2\right)}{m_{EE^c}^2+\lambda_2^2N_2^2}\,,\nonumber\\
R_{2,3}&\simeq&\pm\sqrt{m_{EE^c}^2+\lambda_2^2N_2^2}\,, 
\end{eqnarray}
where we neglect the terms of the order $v_i/N_i$. The associated
eigenvectors with $R_2$ and $R_3$ form a Dirac spinor with mass
$\sqrt{m_{EE^c}^2+\lambda_2^2N_2^2}$.  $R_1$ is the lightest mass
eigenvalue ($\ll m_{\nu\nu^c}$) which represents the lightest mass eigenstate. 
The corresponding eigenvectors can be found in a straightforward manner
under the same assumption that we have used to get the eigenvalues and the
transformation from the mass eigenstates to the flavor eigenstates becomes
\begin{eqnarray}
\displaystyle
\left(\begin{array}{l}
\vspace{0.18cm}
|\nu_L\rangle\\
\vspace{0.18cm}
|N_L\rangle\\
\vspace{0.18cm}
|N_L^c\rangle\\
\vspace{0.18cm}
|\nu_L^c\rangle\\
\vspace{0.18cm}
|S_L^c\rangle\\
\end{array}
\right)= \left(\begin{array}{ccccc}
\vspace{0.18cm}

       0      & \frac{\lambda_2 N_2}{R} & -\frac{\lambda_2 N_2}{R}  & -\frac{m_{EE^c}}{R} &    -\frac{m_{EE^c}}{R}        \\
\vspace{0.18cm}

       0      & \frac{m_{EE^c}}{R}  & -\frac{m_{EE^c}}{R}  & \frac{\lambda_2 N_2}{R}             & \frac{\lambda_2 N_2}{R}\\
\vspace{0.18cm}
       0       & \frac{1}{2}  &       \frac{1}{2}       & 0  & 0  \\
\vspace{0.18cm}

\frac{\lambda_2 N_2}{R}        &      0        & 0 &       \frac{m_{EE^c}}{R}        &    -\frac{m_{EE^c}}{R}         \\
\vspace{0.18cm}

-\frac{m_{EE^c}}{R}     &  0 & 0 &   \frac{\lambda_2 N_2}{R}     &    -\frac{\lambda_2 N_2}{R}   \\ 
\end{array}
\right)\left(\begin{array}{l}
\vspace{0.18cm}
|\nu_1\rangle\\
\vspace{0.18cm}
|\nu_2\rangle\\
\vspace{0.18cm}
|\nu_3\rangle\\
\vspace{0.18cm}
|\nu_4\rangle\\
\vspace{0.18cm}
|\nu_5\rangle\\
\end{array}
\right)\,,\label{ALRmasstoflavor}
\end{eqnarray}
where $R\equiv\sqrt{2(m_{EE^c}^2+\lambda_2^2N_2^2)}$.

At this stage there appears another potential problem in that the lightest mass
eigenstate is
$|\nu_1\rangle=\frac{1}{\sqrt{m_{EE^c}^2+\lambda_2^2N_2^2}}\left[\lambda_2
N_2 |\nu_L^c\rangle -m_{EE^c} |S_L^c\rangle\right]$.  Both $\nu_L^c$ and
$S_L^c$ transform as singlets under the weak interaction gauge group
$SU(2)_L$. This presumed physical neutrino state does not couple with
the left handed SM particles at the low energy scale where the neutrinos
are relevant.\footnote{Even though $S_L^c$ is a part of $SU(2)_{R^{\prime}}$ doublet
and it is possible to consider its interaction with left handed SM leptons
through Higgs bi-doublet at the scales where ALR gauge group is not
broken.} The mass is of the order of magnitude of 
$m_{\nu\nu^c}^2/m_{EE^c}$, which is the expected order of magnitude for 
neutrinos. We thus have two problems:  the 
active neutrinos have a mass of 
the same order of magnitude as the up quark mass, and the lightest 
neutrino is composed of $SU(2)_L$ singlets.   After considering neutrinos in
the Inert model, we will address the above issues and discuss the possible
solutions.

\section{Neutrinos in the Inert Model}\label{sec:Inert}
The neutral fermion mass matrix has similarities with that of the ALR model. 
The Yukawa interactions are invariant under the $SU(2)_I$ group which transforms
$\left(N_L\;E_L\right)\Leftrightarrow\left(\nu_L\;e_L\right),\,
d_L^c\Leftrightarrow h_L^c,$ and $\nu_L^c\Leftrightarrow S_L^c$.  By
following the same procedure as for the ALR symmetric model, one can
obtain the Yukawa Lagrangian for the Inert group and the relevant part of it
reads 
\begin{eqnarray}
{\cal L_Y^{\prime\,\rm 0}}\;=\!\!&&\lambda^{\prime}_1\left[v_1 N_L^c 
S_L^c+v_2\nu_L^cN_L^c+v_3\left(\nu_L\nu_L^c+N_LS_L^c\right)+N_1\left(N_L N_L^c+E_LE_L^c\right)\right.\nonumber\\
\!\!&&\left.+N_2\left(\nu_L N_L^c+e_LE_L^c\right)\right]+\lambda^{\prime}_2\left[ v_1 e_L e_L^c+v_2 e_L^c E_L 
\right]+\lambda^{\prime}_3v_3u_Lu_L^c\,,
\label{Inertyukawa}
\end{eqnarray} 
where the $\lambda^{\prime}_3$-term is especially included to show that, as
 in the ALR case, the $\nu_L$ neutrinos get Dirac masses from the same Higgs scalar as the up quark.  
Without fine tuning between $\lambda^{\prime}_1$ and $\lambda^{\prime}_3$,
the Inert model has the same Dirac mass problem for active
neutrinos as ALR.  The mass matrix for one generation in the basis
$(\nu_L,N_L,N_L^c,\nu_L^c,S_L^c)$ 
\begin{eqnarray}
{\cal M^{\prime}_{\rm neutral}}= \left(\begin{array}{ccccc}
\vspace{0.18cm}
       0      &      0         & \lambda^{\prime}_1 N_2  & \lambda^{\prime}_1 v_3  &     0          \\
\vspace{0.18cm}
       0      &      0         & \lambda^{\prime}_1 N_1  & 0              & \lambda^{\prime}_1 v_3  \\
\vspace{0.18cm}
\lambda^{\prime}_1 N_2 &  \lambda^{\prime}_1 N_1 &       0        & \lambda^{\prime}_1 v_2  & \lambda^{\prime}_1 v_1  \\
\vspace{0.18cm}
\lambda^{\prime}_1 v_3 &      0         & \lambda^{\prime}_1 v_2  &       0        &     0          \\
\vspace{0.18cm}
       0      &  \lambda^{\prime}_1 v_3 & \lambda^{\prime}_1 v_1  &       0        &     0          \\ 
\end{array}
\right)\,.
\end{eqnarray}
Here we recall $\lambda^{\prime}_1 v_1\equiv m^{\prime}_{ee^c},
\lambda^{\prime}_1 N_1\equiv m^{\prime}_{EE^c}$, and $\lambda^{\prime}_1
v_3\equiv m^{\prime}_{\nu\nu^c}$.  The secular equation becomes %
\begin{eqnarray}
\left(R^{\prime}-m^{\prime}_{\nu\nu^c}\right)\left(R^{\prime}+m^{\prime}_{\nu\nu^c}\right)&&\!\!\!\!\big(R^{\prime
3} -R^{\prime} \left(m^{\prime 2}_{EE^c}+\lambda^{\prime 2}_1 (N_2^2+
v_1^2+ v_2^2 )+m^{\prime 2}_{\nu\nu^c}\right)\nonumber\\
&&+2m^{\prime}_{\nu\nu^c}\left(\lambda^{\prime 2}_1 v_2 N_2+
\lambda^{\prime}_1 v_1m^{\prime}_{EE^c}\right)\big)=0\,, \end{eqnarray} %
where there are two eigenvalues $\pm m_{\nu\nu^c}$ which are exact (unlike
the ALR model).  
Diagonalization of the mass matrix gives the following
eigenvalues, under the assumption $v_i\ll N_i$ 
\begin{eqnarray}
R^{\prime}_1&\simeq&-\frac{2m^{\prime}_{\nu\nu^c} \left(\lambda^{\prime}
v_1 m^{\prime}_{EE^c}+\lambda^{\prime 2}_1 v_2
N_2\right)}{m_{EE^c}^2+\lambda^{\prime 2}_1 N_2^2}\,,\nonumber\\
R^{\prime}_{2,3}&\simeq&\pm\sqrt{m_{EE^c}^2+\lambda^{\prime
2}_1N_2^2}\,,\nonumber\\ 
R^{\prime}_{4,5}&=&\pm m^{\prime}_{\nu\nu^c}\,,
\end{eqnarray} 
It is clear from the ALR symmetric model results that there will be two very
heavy neutrinos, one very light neutrino, and two neutrinos with masses of the 
scale of up quark mass. 
The lightest neutrino is $|\nu^{\prime}_1\rangle=\frac{1}{\sqrt{m^{\prime
2}_{EE^c}+\lambda^{\prime 2}_1 N_2^2}}\left(\lambda^{\prime}_1 N_2
|\nu_L^c\rangle -m^{\prime}_{EE^c} |S_L^c\rangle\right)$ and suffers from
the same problem that the ALR symmetric model neutrino does.    We will discuss possible remedies these
problems for both models in the next section.

\section{Solutions to the Neutrino Mass Problem}\label{sec:solution}
As shown in the last two sections, both ALR and Inert models (as
well as the conventional LR symmetric model) have a Dirac neutrino
mass problem at the first stage.  This seems to be a general feature of 
string-inspired low-energy $E_6$ models.  Both models under consideration 
predict that the lightest neutrino state, while having a reasonable mass, is 
composed of $SU(2)_L$
singlets. Furthermore, in their neutral fermion spectrum, there are
neutrino eigenstates having masses of the order of the up quark mass (or the
electron mass for the conventional LR model).  There are three methods
discussed in the literature to rectify this latter neutrino mass
problem.  The smallness of the neutrino masses can be achieved by
introducing a discrete symmetry (the DS method) \cite{Branco:1987yg, Campbell:1986xd, Ma:1986bu, Masiero:1986yf, Masiero:1986ux, 
Ma:1987ab}, or by including a
non-renormalizable higher-order dimensional operators (the HDO method) \cite{Nandi:1985uh, Cvetic:1992ct, Derendinger:1985cv, 
London:1986up}, 
or
using light $E_6$ singlet fields ( the additional neutral fermion (ANF)
method ) \cite{Mohapatra:1986bd, Mohapatra:1986ks, Mohapatra:1986aw}.  We discuss the features
of the models under consideration for each of these three methods. 
As we will see, the predictions are quite different.  The DS method is the
most attractive method among them as it doesn't require any further
particles or the existence of some intermediate scale.  However, it
does not help in non-SUSY framework (at least for the simplest 
discrete symmetry) without introducing many additional particles.  
The HDO method will offer a partial 
solution but {\it does not predict any light sterile
neutrino(s)} and requires new Higgs fields from
$\bf{\overline{27}}$ representation of $E_6$, and the existence of some
intermediate scale, which further breaks the gauge groups of the model. 
The ANF method works well for predicting the lightest state with sterile
neutrino(s) mixing and can explain the LSND result.  However, the method
requires a discrete symmetry as well as new neutral $E_6$ fermion fields, and
a pair of ${\bf{27}+\bf{\overline{27}}}$ split Higgs multiplets whose vev's
do not require hierarchical separation. 

\subsection{The Discrete Symmetry Method}
Following the above discussion, 
the Discrete Symmetry (DS) method is
the most economical.  The symmetry
transformation which is introduced should restrict the existence of
the Dirac mass term $v_3 \nu_L\nu_L^c$ at tree level in the Lagrangian
(Eqs.~(\ref{ALRyukawa}) and (\ref {Inertyukawa})) while allowing 
couplings so that one-loop radiative corrections can be used to generate
naturally small Dirac masses for neutrinos (although it may be necessary 
to put some upper limits for products of some Yukawa couplings).  The symmetry
should also avoid rapid proton decay.

In the supersymmetric versions of both the ALR symmetric and the Inert
model, there exist leptoquark couplings mediated by $h_L$ and $h_L^c$
particles and these couplings are needed to induce nonzero one-loop
neutrino mass.  Since we do not consider the existence of the Higgs fields
carrying $SU(3)$ color, there is no direct analogy in non-SUSY scenarios coming from the
supersymmetrized versions of the models.  It should be noted that the rapid
proton decay is not an issue.

An example of such a symmetry, which was 
considered within the SUSY framework of the general $E_6$ model \cite{Branco:1987yg, Campbell:1986xd} is 
$Z_2\otimes Z_3$.   The $Z_2$ in that case was related to SUSY, and in 
this non-SUSY framework a simple $Z_3$ will suffice.   It is not difficult
to see that such symmetries must be able to differentiate between
generations as long as a non-zero one-loop Dirac neutrino mass is generated
while at the same time eliminating the tree level mass term (see \cite{Masiero:1986ux, Ma:1987ab}
for details).

\begin{figure}[htb]
\begin{center}
\includegraphics[height=1.8in,width=3.5in,angle=0]{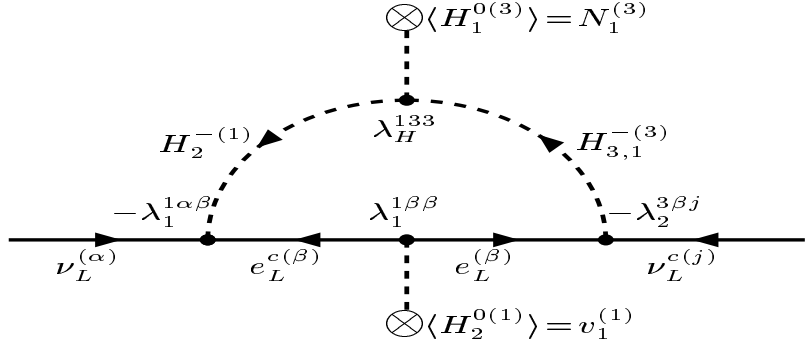}
\caption{The one-loop Dirac masses for $\nu_L^{(\alpha)}\nu_L^{c(j)}$ where $\alpha$ runs over only the second and the third 
generations.}
\label{fig:diracmass}
\end{center}
\end{figure}

In both models considered here, tree level masses of both the neutrinos and
the up quark are obtained from the Higgs field
with vev $v_3$.  As we shall show shortly, eliminating the $v_3$-term will 
cause difficulty.  Let us consider the ALR model.  The Inert model
has very similar features. For a one-loop Dirac neutrino mass, as depicted
in Fig.~{\ref{fig:diracmass}} for a specific choice, the $H_1^0$ $SU(2)_L$ Higgs singlet, $H_2$ and
$H_3$ $SU(2)_L$ Higgs doublets must all participate.  Restating their 
particle content from Eq.~(\ref{higgsALR}) 
\begin{eqnarray}
H_1=\left(H_1^+\;H_1^0\right),\;\; H_2 = \left(\begin{array}{c} H_2^0\\
H_2^- \\ \end{array} \right),\;\; H_{3,1} =\left(\begin{array}{c} H_3^+\\
H_3^0\\ \end{array} \right)\,, \label{higgsALR2} 
\end{eqnarray}  
where $\langle H_2^0\rangle = v_1, \langle H_3^0\rangle = v_3,$ and $\langle
H_1^0\rangle = N_1$.  Here $H_{3,1}$ represents the first column of the
$H_3$ bi-doublet.  Then the relevant terms in the Yukawa Lagrangian
Eq.~(\ref{ALRyukawa}), including the charged Higgs fields interactions
are 
\begin{eqnarray} 
\Delta{\cal L}_{ALR}&=&\lambda_1\left[H_2^0 e_Le_L^c
- H_2^- \nu_Le_L^c - H_1^+ E_L\nu_L +H_1^0 E_L E_L^c\right]\nonumber\\ &&+
\lambda_2 \left[H_3^0 \nu_L\nu_L^c - H_3^+\nu_L^c e_L - H_2^- E_L^c\nu_L^c
\right]\,.\label{ALRDS} 
\end{eqnarray} 

We also need the trilinear Higgs interactions to compute the
diagram given in Fig.~\ref{fig:diracmass}.  The allowed interactions are %
\begin{eqnarray} \Delta{\cal L}_{H}&=&-\lambda_H H_2^T \,\epsilon\, H_3
H_1^0\nonumber\\ &=&\lambda_H H_2^- H_3^+ H_1^0 - \lambda_H H_2^0 H_3^0
H_1^0, \end{eqnarray} %
where $\lambda_H$ is a dimensionful constant. 

Without specifying the charges of the fields under the discrete symmetry, let us
consider the one-loop mass diagram.  One can assign suitable charges to
both Higgs and fermion fields such that the $H_3^0 \nu_L\nu_L^c$ term, a tree
level Dirac mass term for $\nu_L$, is transformed to itself with a nonzero
phase factor and one is then required to set $\lambda_2$ zero for all 3
generations.  If the SM charged leptons and $H_2^-$ and $H_3^-$ fields are
circulating in the loop, the $H_3^+\nu_L^c e_L$ interaction is also proportional to
$\lambda_2$, thus this diagram vanishes.  For the case when $E_L, E_L^c$ are circulating in the loop
instead of the SM charged leptons, it is still necessary to have a nonzero
$\lambda_2$ (clear from Eq.~(\ref{ALRDS})) to get a one loop Dirac mass for
$\nu_L$.  Therefore, eliminating $H_3^0 \nu_L\nu_L^c$ by the $Z_3$ 
symmetry also prevents one-loop mass generation.  This fact
remains true for higher order loops.  The same conclusion applies for the
Inert model as well.

One could consider the possibility that $v_3$ could be zero.  Then
$\lambda_2$ doesn't need to be zero and one-loop Dirac neutrino mass
generation is possible.  In that case, however, all the up quarks ($u,c,t$)
become massless at tree level and generating the top quark mass from a 
loop diagram is very unlikely, within the context of perturbation theory.  

It still is possible to generate a one-loop Dirac neutrino mass if many 
additional fields are introduced.  For example, if one allows for 
``generations'' of Higgs fields, then the $\lambda$ parameters above are all 
third rank tensors.  In such a case, one can arrange the potential
so that some of the $H_3$ vev's vanish.  Then the discrete symmetry can 
couple $\nu_L\nu_L^c$ to fields that do not get vev's, thus allowing a 
one-loop Dirac mass to be generated. To do that,
let's assign the following charges for the matter fields under $Z_3$
\begin{eqnarray}
Z_3:&&\left[Q,d^c_L,h_L,h_L^c,L,\nu_L^c\right]^{(i)}\to \eta \left[Q,d^c_L,h_L,h_L^c,L,\nu_L^c\right]^{(i)}\,,\nonumber\\
&&F_1^{(1)}\to \eta^{-1}F_1^{(1)}\,,\; F_1^{(2)}\to F_1^{(2)}\,,\; F_1^{(3)}\to \eta F_1^{(3)}\,,\nonumber\\
&&H^{(1)}\to \eta^{-1}H^{(1)}\,,\; H^{(2)}\to \eta H^{(2)}\,,\; H^{(3)}\to H^{(3)}\,,\nonumber\\
&&S_L^{c(1)}\to \eta^{-1}S_L^{c(1)}\,,\; S_L^{c(2)}\to \eta S_L^{c(2)}\,,\; S_L^{c(3)}\to S_L^{c(3)}\,,
\end{eqnarray}
where $F_1$ is the first column of the bidoublet $F$, and similarly the Higgs fields as
\begin{eqnarray}
Z_3:&&H_{3,1}^{(1)}\to \eta^{-1}H_{3,1}^{(1)}\,,\; H_{3,1}^{(2)}\to H_{3,1}^{(2)}\,,\; H_{3,1}^{(3)}\to \eta H_{3,1}^{(3)}\,,\nonumber\\
&&H_2^{(1)}\to \eta^{-1}H_2^{(1)}\,,\; H_2^{(2)}\to \eta H_2^{(2)}\,,\; H_2^{(3)}\to H_2^{(3)}\,,\nonumber\\
&&H_1^{0(1)}\to \eta^{-1}H_1^{0(1)}\,,\; H_1^{0(2)}\to \eta H_1^{0(2)}\,,\; H_1^{0(3)}\to H_1^{0(3)}\,,\nonumber\\
&&H_{3,2}^{(i)}\to \eta H_{3,2}^{(i)}\,,\;H_S^{(i)}\to \eta H_S^{(i)}\,,
\end{eqnarray}
where the rest of the fields are assumed to be invariant under $Z_3$ and $\eta^3=1$. In this particular choice we take
the vev of $H_{3,1}, v_3^{(3)}$, as zero. Then, the Lagrangian for the ALR symmetric model, given in Eq.~(\ref{YukwALR})
reduces to
\begin{eqnarray}
{\cal 
L_Y}=&&\!\!\!-\lambda_1^{1\alpha\beta}\left[H_2^{(1)}L^{(\alpha)}e^{c(\beta)}_L+H^{(1)}H_{3,2}^{(\alpha)}e^{c(\beta)}_L+H^{(1)}L^{(\alpha)}H_1^{+(\beta)}+H_2^{(1)}F_1^{(\alpha)}S^{c(\beta)}_L\right. 
\nonumber\\
&&\left. +H^{(1)}H_{3,1}^{(\alpha)}S^{c(\beta)}_L+H^{(1)}F_1^{(\alpha)}H_1^{0(\beta)} \right]+\lambda_2^{3ij}\left[H_{3,1}^{(3)}L^{(i)}\nu_L^{c(j)}+F_1^{(3)}H_{3,2}^{(i)}\nu_L^{c(j)}\right.\nonumber\\
&&\left.+F_1^{(3)}L^{(i)}H_S^{(j)}\right]+\lambda_3^{1ij}\left[H_{3,1}^{(1)}u^{c(i)}_L Q^{(j)}+H_{3,2}^{(1)}h^{c(i)}_L Q^{(j)}\right]+\lambda_4^{2ij}H_2^{(2)}Q^{(i)} d^{c(j)}_L\nonumber\\
&&+\lambda_5^{2ij}H_1^{(2)}h^{(i)}_L h^{c(j)}_L + \lambda_6^{ijk}h^{(i)}_Ld^{c{(j)}}_LH_S^{(k)}\,,
\label{YukALRZ3}
\end{eqnarray}
where $\alpha$ and  $\beta$ run only over the second and third generations. 
Now, the only tree level Dirac mass term for $\nu_L$, $\lambda_2^{3ij}H_{3,1}^{(3)}L^{(i)}\nu_L^{c(j)}$, 
vanishes if all the particles are neutral due to zero vev $v_3^{(3)}$. 
Note that writing the Lagrangian for the Inert model can be done easily by applying the following 
substitutions to Eq.~(\ref{YukALRZ3}); $F_1\Leftrightarrow L^{\prime},\, u^c_L\Leftrightarrow d_L^c,\, e^c\Leftrightarrow \nu^c,\, 
H_{3,1}\Leftrightarrow H_2,\, H_1\rightarrow H_D,\, H_S\rightarrow 0$. The grouping of the terms in Inert case will be slightly 
different. We will stick the ALR case in the rest of the subsection.

Due to the radiative corrections based on the remaining interactions given in Eq.~(\ref{YukALRZ3}), 
$\nu_L^{(i)}\nu_L^{c(j)}$ Dirac masses are induced through one-loop diagram shown in Fig.~\ref{fig:diracmass}. 
If we assume that the product $\lambda_H N_1$ is of the same order as the charged Higgs masses, which 
are further assumed degenerate and much heavier than any fermion in the loop, the magnitudes of the 
Dirac masses are roughly estimated as
\begin{eqnarray}
M^{\alpha j}_{\nu\nu^c}=\frac{m_{\tau\tau^c}}{16 \pi^2}\lambda_1^{1\alpha 3}\lambda_2^{33j}.
\end{eqnarray}
In order for such radiative masses to be of the order of $10^{-1}$ eV, the product of the relevant Yukawa couplings   
$\lambda_1^{1\alpha 3}\lambda_2^{33j}$ should be less than $O(10^{-8})$. It is further possible to generate very light 
Majorana masses for both $S^c_L$ and $\nu^c_L$ through one-loop.\footnote{Neither $N_L$ nor $N_L^c$ can get such one-loop Majorana 
masses in this framework.}  Majorana masses for $S^c_L$ are obtained by replacing the tau lepton in 
Fig.~\ref{fig:diracmass} with the $E$ lepton, but are very supressed ($\sim\lambda_1^2m_{H^-}^2/m_{EE^c}$) and similarly for $\nu^c_L$. 
If we include these Majorana masses, this opens up the possibility of having so-called pseudo-Dirac neutrinos when $M_{S^cS^c}, M_{\nu^c\nu^c}\ll M_{\nu\nu^c}$ is satisfied \cite{Langacker:2004xy}. 

Such models have far too many parameters 
to be predictive and are very contrived. We thus turn to the HDO and 
ANF schemes, which are much more predictive.

\subsection{The HDO Method in the ALR and the Inert Models}\label{subsection:HDO}
This method has been discussed in the framework of rank-6 version of the LR
symmetric model \cite{Nandi:1985uh} where it has been shown that the higher dimensional 
operators (HDO), specifically
dimension-5 operators, give sizable contributions to the neutral sector of
the fermion mass matrix.  The method requires the existence of an intermediate scale
at which the group is broken to the SM gauge group.  Two
of the Higgs fields (for our discussion, $H_1$ and $H_S$ in the ALR case,
and $H_D$ in the Inert case) will acquire vev's of the order of the
intermediate scale ($\sim 10^{11}$ GeV).  

The leading HDO Yukawa interactions are the dimension-5
operators. If we neglect the contributions coming from operators with $\rm dim
> 5$,\footnote{It is safe to neglect them since they are suppressed by some
quadratic, cubic or higher powers of the compactification scale, $M_c$($\sim
10^{18}$ GeV).} the non-renormalizable
dimension-5 operator is
\begin{eqnarray}
{\cal L^{\rm (5)}_Y}=\frac{f}{M_c}\psi^T({\bf{27}})\,\epsilon\,H(\overline{{\bf{27}}})\,C\, 
H^T(\overline{{\bf{27}}})\,\epsilon\,\psi({\bf{27}})\,,
\label{dim5Lag}
\end{eqnarray}
where the Higgs fields $H$ are from the $\overline{\bf{27}}$ representation
of $E_6$ and their quantum numbers are taken as the opposite of the
ones listed in Table \ref{tablehiggs}.  Here, $M_c$ is the  
compactification scale, or $10^{18}$ GeV.  The inclusion of the above
dimension-5 interactions will modify all entries in the fermion
sector (both the charged and the neutral fields).  However, from Table
\ref{tablepartic}, it is possible to show that except the $\nu_L^c-S_L^c$
submatrix in the neutral sector all entries get contributions which are
negligible compared to with their dimension-4
entries.\footnote{Negligible contributions are either 0, or $\frac{f v_i
v_j}{M_c}$, or $\frac{f v_i N_j}{M_c}$ form to the appropriate entries, but 
not $\frac{f N_i N_j}{M_c}$.}

The $\nu_L^c-S_L^c$ submatrix, a null $2\times 2$ matrix at the dimension-4 level, becomes, in the ALR model 
\begin{eqnarray}
{\cal M}_{\nu^c-S^c}=\left(\begin{array}{cc}
K_1  & K_{12}\\
K_{12} & K_2
\end{array}
\right)\,,
\label{vSsubmatrix}
\end{eqnarray}
where $K_{12}\equiv 2 f \frac{N_1 N_2}{M_c}$ and $K_{i}\equiv
f\frac{N_i^2}{M_c}$.  Obviously, $K_i\sim K_{12}\simeq 10^4$ GeV for an
intermediate scale $10^{11}$ GeV and the coupling constant $f$ is of order
of unity. The nonzero $2\times 2$ submatrix with large entries 
gives a new ``see-saw-like" structure to the $5\times 5$ matrix. 
The submatrix in the $(\nu_L^c, S_L^c)$ basis will induce  small but non-zero
entries in the upper-left $2\times 2$ submatrix spanned by $(\nu_L, N_L)$. 
The mass eigenvalues for the matrix in Eq.~(\ref{MatrixALR}) with the above
modification become
\begin{eqnarray}
R_1&\simeq&\frac{(\lambda_1\lambda_2 v_3\sqrt{K_1}N_2+m_{\nu\nu^c}M_{EE^c}\sqrt{K_2})^2+\lambda_1\lambda_2 v_3 N_2 m_{\nu\nu^c} m_{EE^c} 
K_{12}}{(m_{EE^c}^2+\lambda_2^2N_2^2)(K_{12}^2-K_1K_2)}\,,\nonumber\\
R_{2,3}&\simeq& \pm\sqrt{m_{EE^c}^2+\lambda_2^2 N_2^2}\,,\nonumber \\
R_{4,5}&\simeq&\frac{1}{2}\left[K_1+K_2\pm\sqrt{(K_1-K_2)^2+4K_{12}^2}\right]\,,
\label{dim5massALR}
\end{eqnarray}
where we use the assumptions $v_i\ll K_i\sim K_{12}\ll m_{EE^c}\sim\lambda_2
N_2$ and neglect all the $m_i^2$ terms.  The first apparent modification
from the mass eigenvalues is that the states with masses $R_{4,5}$, which 
previously had masses of the order of the up quark mass, now get modified at the
scale $K_{1,2,12}\sim 10^4$ GeV. After the diagonalization, the
transformation matrix (the analogous to the dimension-4 case (Eq. 
(\ref{ALRmasstoflavor})) in the dimension-5 level) is
\begin{eqnarray}
\displaystyle
\left(\begin{array}{l}
\vspace{0.20cm}
|\nu_L\rangle\\
\vspace{0.20cm}
|N_L\rangle\\
\vspace{0.20cm}
|N_L^c\rangle\\
\vspace{0.20cm}
|\nu_L^c\rangle\\
\vspace{0.20cm}
|S_L^c\rangle\\
\end{array}
\right)= \left(\begin{array}{ccccc}
\vspace{0.18cm}

a_1 m_{EE^c}        &      0      & 	0	  &\frac{a_1 \lambda_2 N_2}{\sqrt{2}} & \frac{a_1 \lambda_2 N_2}{\sqrt{2}}       \\
\vspace{0.18cm}

-a_1 \lambda_2 N_2   &      0      & 	0	  &\frac{a_1 m_{EE^c}}{\sqrt{2}}       &\frac{a_1 m_{EE^c}}{\sqrt{2}}\\
\vspace{0.18cm}
       0            &      0      &     	0	  &          \frac{1}{\sqrt{2}}        & -\frac{1}{\sqrt{2}} \\
\vspace{0.18cm}

       0            &  a_2 K_{12} &    a_3K_{12}  &       		0		         &   0       \\
\vspace{0.18cm}

       0       &   a_2 (R_4-K_1)  & a_3 (R_5-K_1) & 		      0	    	         &    0   \\ 
\end{array}
\right)\left(\begin{array}{l}
\vspace{0.20cm}
|\nu_1\rangle\\
\vspace{0.20cm}
|\nu_2\rangle\\
\vspace{0.20cm}
|\nu_3\rangle\\
\vspace{0.20cm}
|\nu_4\rangle\\
\vspace{0.20cm}
|\nu_5\rangle\\
\end{array}
\right)\,,\label{ALRmasstoflavor5}
\end{eqnarray}
where $a_1\equiv\frac{1}{\sqrt{m_{EE^c}^2+\lambda_2N_2^2}}, a_2\equiv
\frac{1}{\sqrt{K_{12}^2+(R_4-K_1)^2}}, a_3\equiv
\frac{1}{\sqrt{K_{12}^2+(R_5-K_1)^2}}$.  The above matrix elements are
derived in the same limit as we used before to get the mass eigenvalues. 
Now, the spectrum consists of one light state, $\nu_1$, and four heavy
states, $\nu_{2,3,4,5}$.  Moreover, the light state is formed by the flavor
states $\nu_L$ and $N_L$ of the form
\begin{eqnarray}
\nu_1\simeq\frac{1}{\sqrt{m_{EE^c}^2+\lambda_2^2 N_2^2}}\left[m_{EE^c}|\nu_L\rangle - \lambda_2 N_2 |N_L\rangle\right]\,,
\end{eqnarray}
which is an acceptable physical state as both $\nu_L$ and $N_L$ are members
of two different $SU(2)_L$ doublets.  Therefore our physical neutrino state
can now couple with the electron and the other SM particles in a desired
way.  The mass of the state is still as light as
$m_{\nu\nu^c}^2/K_{1,2,12}\;($or$\,(\lambda_1 v_3)^2/K_{1,2,12})\sim 0.02$
eV when we take the $m_{\nu\nu^c}$ around the mass of the up quark.

One can repeat the same calculation for the Inert model.  The features are
very similar.  Except the $\nu^c_L-S_L^c$ submatrix, all other entries get
negligible contributions from Eq.~(\ref{dim5Lag}) and in the submatrix, the
corresponding $SU(2)_I$ Higgs doublet $H_D$ from the $\overline{\bf
27}$-plet of $E_6$ is involved and the submatrix will be the
same as the one in Eq.~(\ref{vSsubmatrix}).  The eigenvalues are slightly
different from the ones given in Eq.~(\ref{dim5massALR})
\begin{eqnarray}
R^{\prime}_1&\simeq&\frac{m_{\nu\nu^c}^{\prime 2}(\lambda^{\prime 2}_1 N_2K_1+m_{EE^c}^{\prime 2} K_2 + 2 \lambda_1^{\prime}N_2 m_{EE^c}^{\prime}K_{12})}{(m_{EE^c}^{\prime 2}+\lambda_2^{\prime 2} N_2^2)(K_{12}^2-K_1K_2)}\,,\nonumber\\
R^{\prime}_{2,3}&\simeq& \pm\sqrt{m_{EE^c}^{\prime 2}+\lambda_2^{\prime 2} N_2^2}\,,\nonumber \\
R_{4,5}&\simeq&\frac{1}{2}\left[K_1+K_2\pm\sqrt{(K_1-K_2)^2+4K_{12}^2}\right]\,,
\label{dim5massInert}
\end{eqnarray}
under the same assumptions  as previously stated. 
Then the transformation matrix can be formed by finding the corresponding
mass eigenstates and it has the same form as the one in the ALR
model given in Eq.~(\ref{ALRmasstoflavor5}).  Note that the results 
differ from each other when we, for example, keep terms in the
$O(v_i/N_i,v_i/K_{1,2,12})$ order.  The lightest state $\nu^{\prime}_1$ is
composed of $\nu_L$ and $N_L$ of the form
\begin{eqnarray}
\nu^{\prime}_1\simeq\frac{1}{\sqrt{m^{\prime 2}_{EE^c}+\lambda^{\prime 2}_1 N_2^2}}\left[m^{\prime}_{EE^c}|\nu_L\rangle - \lambda^{\prime}_1 N_2 |N_L\rangle\right],
\end{eqnarray}
where the flavor states $\nu_L$ and $N_L$ mix, like in the ALR model.
From these results we see that the HDO method solves the
problems in both models, under the assumption that there exists
an intermediate scale at the order of $10^{11}$ GeV and both $N_1$ and
$N_2$ get vev's at that scale.

Since there is only one light state (per generation, of course), there is 
no sterile neutrino in the model.  The $N_L$ only couples to the $E$, and 
which is very heavy, the net effect of the mixing (in either the ALR or 
Inert model) will be to lower the coupling of the electron neutrino to the 
electron and $W_L$-boson.  For the ALR case (the Inert case is basically the 
same), the coupling is reduced by a factor of 
${\lambda_1N_1\over\sqrt{\lambda_1^2N_1^2+\lambda_2^2N_2^2}}$.  Since the 
mixing must be small, $\lambda_2N_2 \ll  \lambda_1N_1$, and this factor then 
becomes $1-{\lambda_2^2N_2^2\over 2\lambda_1^2N_1^2}$.

This reduction would give a very clear signature for the model.  The 
electron neutrino would not oscillate into a sterile neutrino (ignoring 
inter-generational mixing), and yet its coupling is reduced relative to the 
standard model.  Similar reductions would occur for the muon and tau 
neutrino interactions. The phenomenological implications of this reduction 
will be discussed in the next Section.

\subsection{The Additional Neutral Fermion Method}
In some $E_{6}$-based superstring-based models, such as those 
with  Calabi-Yau compactification, in addition to 
the $\bf{27}$ and $\bf{\overline{27}}$ representations of  $E_6$ for 
the matter multiplets, there typically exist split multiplets, parts 
of 
the $\bf{27}+\bf{\overline{27}}$ representations, as well as
some $E_6$ singlets 
$\bf{1}$ \cite{Witten:1985bz, Doi:1987nk, Albright:1986rv}.  We have already considered the existence of such Higgs 
multiplets by considering  $\bf{\overline{27}}$ components of the 
above  
$\bf{27}+\bf{\overline{27}}$ representation inducing dimension-5 
terms (of the form discussed in the 
previous subsection).  In addition to the $(\bf{27})^3$ and the 
higher-dimensional
$(\bf{27}\cdot\bf{\overline{27}})^2$ types of interactions, we may 
have $\bf{27}\cdot\bf{\overline{27}}\cdot\bf{1}$ type of interactions 
as well. The Additional Neutral Fermion 
(ANF) method follows this approach. The existence of $E_6$ singlets 
(and thus the $\bf{27}\cdot\bf{\overline{27}}\cdot\bf{1}$ 
interactions) has been discussed in different context of the 
superstring models \cite{Mohapatra:1986bd, Mohapatra:1986ks, Mohapatra:1986aw} to tackle 
the rapid proton decay problem, 
large neutrino mass problem and others. In order to give light
neutrino masses consistent 
with present experimental observations, the additional Higgs fields are
required to have vev's chose in a 
strong hierarchical way, which seems unnatural. 
Such an odd pattern, however, is not necessary 
in the non-SUSY versions of the models discussed here. 
We discuss the method in the ALR symmetric model and 
later point out the difference with the Inert 
model. 

In the ALR model,  we consider one additional $E_6$ neutral 
fermion singlet\footnote{For simplicity, we assume one additional 
field 
$\phi$ even when we extend our discussion to the three generation 
case later in this section.} $\phi$, and one pair of 
$\bf{27}+\bf{\overline{27}}$ Higgs multiplets $H + \overline{H}$ (the 
Betti-Hodge number $b_{1,1}=1$).  We do not include a 
corresponding $\bf{27}+\bf{\overline{27}}$ chiral fermion multiplet 
relevant for supersymmetrized versions of the models 
considered in future studies.\footnote{In principle, one 
can add such new fields and the corresponding interactions. We would 
like to be as  conservative as possible as far as the number of 
new parameters are concerned.} Let us assume that both $H$ and 
$\overline{H}$ have $\nu^c$-like and $S^c$-like components 
$H_{\nu^c,S^c}, \overline{H}_{\nu^c,S^c}$. 
Since we don't want to alter the interactions in the $(\bf{27})^3$ 
sector discussed earlier, we assume that only 
$\overline{H}_{\nu^c,S^c}$ get nonzero vev's and further, that there is 
 a $Z_2$ discrete symmetry under which all fields except 
$\phi,H_{\nu^c,S^c}$ and $\overline{H}_{\nu^c,S^c}$ have even 
charges. Therefore, two additional gauge invariant interactions for 
one generation  survive of the form
\begin{eqnarray}
\Delta{\cal L}^{\phi}_{ALR} = \lambda_S S^c_L \overline{H}_{S^c}\phi 
+ \lambda_{\nu} \nu^c_L \overline{H}_{\nu^c}\phi\,.
\end{eqnarray}
Then, the mass 
matrix in the neutral fermion sector in the ($\nu_L, N_L, N^c_L, 
\nu^c_L, S^c_L, \phi$) basis can be obtained directly by adding a 
column and a row for $\phi$ field to the one given in 
Eq.~(\ref{MatrixALR}) 
\begin{eqnarray}
{\cal M}_{\rm neutral}= \left(\begin{array}{cccccc}
\vspace{0.18cm}
       0      &      0         & \lambda_2 N_2  & \lambda_2 v_3  
&     0         & 0  \\
\vspace{0.18cm}
       0      &      0         & -\lambda_1 N_1  & 0              & 
-\lambda_1 v_3 & 0 \\
\vspace{0.18cm}
\lambda_2 N_2 &  -\lambda_1 N_1 &       0        & \lambda_2 v_2  & 
-\lambda_1 v_1 & 0 \\
\vspace{0.18cm}
\lambda_2 v_3 &      0         & \lambda_2 v_2  &       0        
&     0         & \lambda_{\nu} V \\ 
\vspace{0.18cm}
      0      &  -\lambda_1 v_3 & -\lambda_1 v_1  &       0        &     
0          & \lambda_{S} \mu \\ 
\vspace{0.18cm}
      0      &       0        &        0        &     \lambda_{\nu} 
V  & \lambda_{S} \mu & 0 \\
\end{array}
\right)\,,\label{MatrixALRANF}
\end{eqnarray}
where we define $\langle \overline{H}_{S^c} \rangle \equiv \mu$ and 
$\langle \overline{H}_{\nu^c} \rangle \equiv V$.

The eigenvalues can be found by following the same methodology as 
before and under the assumption $v_i, m_{ee^c}, m_{\nu\nu^c} \ll  
N_1, N_2, \mu, V$ (we assume $N_i\sim \mu, V$) giving
\begin{eqnarray}
R_{1,2}&\simeq&\pm\frac{m_{\nu\nu^c}m_{ee^c}(\lambda_2 N_2)(\lambda_S 
\mu)(\lambda_{\nu} V)}{\left(\lambda_2^2 
N_2^2+m_{EE^c}^2\right)\left(\lambda_S^2 \mu^2 + \lambda_{\nu}^2 V^2 
\right)}\,,\nonumber\\
R_{3,4}&\simeq&\pm\sqrt{\lambda_S^2 \mu^2 + \lambda_{\nu}^2 
V^2}\,,\nonumber\\
R_{5,6}&\simeq&\pm\sqrt{\lambda_2^2 N_2^2+m_{EE^c}^2}\,.
\label{eigenvaluesANF}
\end{eqnarray}
Now, we have two light eigenvalues $R_{1,2}$. The masses of these states can be
approximated as $(m_{\nu\nu^c} m_{ee^c})/m_{EE^c}$ and could possibly be 
in the experimentally favored region while obeying 
the the experimental bounds on $\nu_L - N_L$ mixing. It is straightforward to 
get the mass eigenstates corresponding to the above eigenvalues. The 
transformation matrix equation from mass to flavor eigenstates is given by
\begin{eqnarray}
\displaystyle
\left(
\begin{array}{l}
\vspace{0.28cm}
|\nu_L\rangle\\
\vspace{0.25cm}
|N_L\rangle\\
\vspace{0.25cm}
|N_L^c\rangle\\
\vspace{0.25cm}
|\nu_L^c\rangle\\
\vspace{0.19cm}
|S_L^c\rangle\\
|\phi\rangle\\
\end{array}
\right)= \left(\begin{array}{cccccc}
\vspace{0.18cm}

\frac{m_{EE^c}\,\cos\theta}{R_5}            & \frac{m_{EE^c}\, 
\sin\theta}{R_5} & 0  & 0 &  \frac{1}{\sqrt{2}}\frac{\lambda_2 
N_2}{R_5}  & \frac{1}{\sqrt{2}}\frac{\lambda_2 N_2}{R_5}      \\
\vspace{0.18cm}

\frac{-\lambda_2 N_2\,\cos\theta}{R_5}        & \frac{-\lambda_2 
N_2\, \sin\theta}{R_5}  & 0  & 0             & 
\frac{1}{\sqrt{2}}\frac{m_{EE^c}}{R_5} & 
\frac{1}{\sqrt{2}}\frac{m_{EE^c}}{R_5}\\
\vspace{0.18cm}
       0       & 0 &       0      & 0  & \frac{1}{\sqrt{2}} & 
\frac{-1}{\sqrt{2}} \\
\vspace{0.18cm}

\frac{-\lambda_S\,\mu\,\sin\theta}{R_3}        & 
\frac{\lambda_S\,\mu\,\cos\theta}{R_3}       & 
\frac{1}{\sqrt{2}}\frac{\lambda_{\nu} V}{R_3} 
&\frac{1}{\sqrt{2}}\frac{\lambda_{\nu} V}{R_3}      &   0 & 0        
\\
\vspace{0.18cm}

\frac{\lambda_{\nu} V\,\sin\theta}{R_3}     & \frac{-\lambda_{\nu} 
V\,\cos\theta}{R_3}  & \frac{1}{\sqrt{2}}\frac{\lambda_S\, \mu}{R_3} 
&  \frac{1}{\sqrt{2}}\frac{\lambda_S\, \mu}{R_3}    &   0 & 0   \\ 
 0                                         &  0 & 
\frac{1}{\sqrt{2}}&\frac{-1}{\sqrt{2}}& 0& 0 \\ 
\end{array}
\right)\left(\begin{array}{l}
\vspace{0.28cm}
|\nu_1\rangle\\
\vspace{0.25cm}
|\nu_2\rangle\\
\vspace{0.25cm}
|\nu_3\rangle\\
\vspace{0.25cm}
|\nu_4\rangle\\
\vspace{0.19cm}
|\nu_5\rangle\\
|\nu_6\rangle\\
\end{array}
\right)\,,\label{ALRmasstoflavorANF}
\end{eqnarray}
where $R_3$ and $R_5$ are given in Eq.~(\ref{eigenvaluesANF}). 
The parameter $\theta$ is arbitrary in the model, but it would be 
fixed both by the requirement that the coupling of $W_L$ to neutrinos and 
leptons must be in agreement with the experimental data and by the 
required mixing angle between active and sterile 
neutrinos. The mass eigenstates $|\nu_3\rangle, 
|\nu_4\rangle, |\nu_5\rangle$, and $|\nu_6\rangle$ corresponding to eigenvalues
$R_{3,4,5,6}$ respectively are heavy and irrelevant to our discussion 
at low energies. There are two light mass eigenstates of the form 
\begin{eqnarray}
|\nu_1\rangle &=& \cos\theta\left(\frac{m_{EE^c}}{R_5}|\nu_L\rangle 
- \frac{\lambda_2 N_2}{R_5} |N_L \rangle \right) + 
\sin\theta\left(\frac{\lambda_{\nu} V}{R_3}|S_L^c\rangle - 
\frac{\lambda_S\,\mu }{R_3} |\nu_L^c \rangle \right)\,,\nonumber\\
|\nu_2\rangle &=& \sin\theta\left(\frac{m_{EE^c}}{R_5}|\nu_L\rangle 
- \frac{\lambda_2 N_2}{R_5} |N_L \rangle \right) - 
\cos\theta\left(\frac{\lambda_{\nu} V}{R_3}|S_L^c\rangle - 
\frac{\lambda_S\,\mu }{R_3} |\nu_L^c \rangle \right)\,.
\label{lightANF}
\end{eqnarray}
The above results apply to the Inert group, with an additional 
constraint coming from $SU(2)_I$ symmetry. Since $\nu_L$ and $S_L^c$ 
form an $SU(2)_I$ doublet, the couplings $\lambda_{\nu}$ and 
$\lambda_S$ are required to be equal.  

Thus, we have two interesting features of the model.  The slight 
suppression of
the coupling of the active neutrino discusssed in the last subsection 
is present.  However, now 
we {\it also} have a sterile neutrino with an arbitrary mixing angle 
with the active neutrino.  This model could then easily accommodate the LSND result 
(if confirmed by MiniBooNE).

With the addition of only one singlet, for simplicity, there are three active neutrinos. 
In this case, $\lambda_{S}$ and 
$\lambda_{\nu}$ have generation indices.   Each active neutrino has 
a light mass, and will mix with an arbitrary mixing angle with 
the sterile neutrino.   Note that in the single-generation case, the 
two light mass eigenstates are, to leading order, identical.  Thus, 
if 
the mixing angle is small for two of the three generations, we will 
have a ${\bf 2+2}$ structure, whereas if it is sizeable for all three 
generations, there will be a ${\bf 3+1}$ structure.  Of course, one 
could introduce several singlet fields, giving more complicated 
structures. 

\section{Discussion of the results}\label{sec:results}
If the LSND result is confirmed by MiniBooNE, the existence of
sterile neutrino(s) at low energies might be unavoidable. Thus is it
important to analyze extensions of the
Standard Model which predict the existence of extra  
neutral fermions, and verify that they have the desired experimental
features. Though we have explicitly considered here the $E_6$
subgroups,
$SU(3)_C\otimes SU(2)_L\otimes SU(2)_{R^{\prime}}\otimes U(1)_{V}$   
(ALR) and $SU(3)_C\otimes SU(2)_L\otimes SU(2)_I\otimes
U(1)_{Y}$ (Inert), and concentrated on the neutrino spectrum in
non-SUSY framework, our work is valid for the $SU(3)_C\otimes
SU(2)_L\otimes SU(2)_R\otimes U(1)_{V}$ (LR) group as well.

These models predict several exotic neutral fermions. We have shown that
both the ALR and Inert models generally predict neutrino sectors
inconsistent with current observations. The lightest state turns out
to contain only $SU(2)_L$ singlets ($\nu_L^c$ and $S_L^c$) which do 
not interact with SM particles. Additionally, in contradiction with
present experimental observations, two more light neutrino states
with masses around the up
quark mass exist. The main reason for such a spectrum is the existence
of tree level Dirac mass term in the Lagrangian. We have discussed  
three possible remedies to this problem.

The most attractive one is the Discrete Symmetry (DS) method which
only requires imposing an extra symmetry. The aim is to eliminate the
tree level Dirac
mass term by assigning suitable charges to the fields under some
discrete symmetries, and generate Dirac neutrino masses through
radiative corrections. The discrete symmetry needs to distinguish
generations. As discussed earlier, there is no way to induce a non-zero
one-loop Dirac mass while eliminating the tree level term. The only  
way out is to have a $SU(2)_L$ Higgs doublet (a part of the  
bidoublet)
with vanishing vev. For this, we considered the simplest symmetry,  
$Z_3$. It leads to Dirac masses from one-loop diagrams which are   
estimated
around $10^{-1}$ eV, by imposing an upper bound to the product of the
Yukawa couplings of the order of $10^{-8}$. It
is also possible to generate very light Majorana masses for $S^c_L$ 
and $\nu_L^c$. Since these masses are much smaller than the Dirac
mass term for $\nu_L$, a spectrum with pseudo-Dirac neutrinos is obtained.

The Higher Dimensional Operators (HDO), the second method, requires
additional Higgs fields from ${\bf{27}}$-plet of $E_6$ and the
existence of some intermediate scale. We introduce interactions which are suppressed by
 one power of the compactification scale, 
through dimension-5 operators. The method solves  the mass problems but does
not predict any sterile neutrino component(s) in the lightest neutrino
state, which is an admixture of $\nu_L$ and $N_L$. The effect of the mixing will 
be to lower the electron neutrino coupling to the electron and the $W_L$ 
boson by a factor of $1- {1\over 2}\Delta_e^2$, where $\Delta_e =\lambda_2 N_2/\lambda_1 N_1$.  
The reduction for the muon and tau neutrino interactions will be given by the same expression, with 
$\Delta_e$ being replaced by $\Delta_\mu$ and $\Delta_\tau$ (which depend on different 
$\lambda_i$ and  $N_i$).   The phenomenological implications are interesting.  If the $\Delta_i$ are 
different, then $e-\mu-\tau$ universality will be violated in neutrino interactions.  By comparing the 
muon decay rate and the rate for leptonic tau decays, one finds \cite{Eidelman:2004wy,PichZardoya:1999yv} that the 
reductions of $1-{1\over 2}\Delta_i^2$ cannot differ by more than $0.005$.  Even if the $\Delta_i$ are 
all the same, however, one would still find a discrepancy in, for example, $\tau\rightarrow\pi\nu_\tau$ 
vs. $\tau\rightarrow\mu\overline{\nu}_\mu\nu_\tau$, which would depend on $\Delta_\mu$, with a 
similar dependence on the electronic decay.  Comparing all of these bounds, we find that none of the 
reductions can exceed $0.005$, leading to a bound, for each generation, of $\lambda_2N_2/
\lambda_1N_1 < 0.1$, which is not particularly fine-tuned.   A more detailed study comparing many 
hadronic decays with the leptonic decays of the $\tau$ could lead to a somewhat more precise bound 
(or, better yet, an indication of a discrepancy).

The last method we have discussed is the Additional Neutral Fermion
(ANF), which requires the existence of both new particles and discrete
symmetries. If one considers an $E_6$ singlet field, the additional  
interactions  will be of the type
${\bf{27}}\cdot {\bf{\overline{27}}}\cdot {\bf{1}}$,  which further   
require additional Higgs doublets from the  
${\bf{27}}+{\bf{\overline{27}}}$ representation. In order not to
alter already existing couplings, the vev's of the new fields need to
be chosen suitably, together with an additional $Z_2$ symmetry. Under
these circumstances we obtain  two light states given in
Eq.~(\ref{lightANF}). The neutrino states have an active neutrino
part of exactly the form predicted by the HDO method, but this time
they mix
with a sterile flavor state (formed by $\nu_L^c$ and $S_L^c$). The   
mixing is completely arbitrary.
If we extend the picture to three generations, the model contains two
structures, ${\bf{2+2}}$ and ${\bf{3+1}}$, which have been
discussed extensively in the literature \cite{Barger:2003qi}. When
the above mixing is sizable only for one generation, only the
${\bf{2+2}}$ structure
arises naturally, since the states in Eq.~(\ref{lightANF}) are
degenerate in the leading order. Otherwise, ${\bf{3+1}}$ is possible.
More
realistically, when we include three generations of  $\nu_L^c$ and    
$S_L^c$, we obtain a ${\bf{3+3}}$ structure.

 Recent analyses show that neither  ${\bf{2+2}}$ nor  ${\bf{3+1}}$  
provide a good
description of the combined  atmospheric, solar, reactor, and
accelerator data even though it appears that ${\bf{3+1}}$ works  
better.
However, there is no consensus about whether the scenarios with four
neutrinos are ruled out or not \cite{Barger:2003qi, Paes:2002ah}. From our 
considerations, the ANF method
allows both
  ${\bf{3+2}}$ or ${\bf{3+3}}$ structures, which enhance the effects 
in favor of LSND
data \cite{Paes:2002ah}.

\begin{acknowledgments}
M.F. and I.T. thank NSERC of Canada for support under grant number 0105354, and  M.S. thanks the National Science Foundation 
under grant PHY-0243400.
\end{acknowledgments}

       
\end{document}